\documentclass[aps,prl,twocolumn,superscriptaddress,showpacs]{revtex4}
\usepackage{latexsym} 
\usepackage{graphicx}
\usepackage{amsmath}

\begin{document}

\title{Divergence at low bias and down-mixing of the current noise in a diffusive superconductor-normal metal-superconductor junction}

\author{E. Lhotel}
\affiliation{CEA-Grenoble, DRFMC / SPSMS / LaTEQS, 17 rue des Martyrs, F-38054 Grenoble cedex 09, France}
\author{O. Coupiac}
\affiliation{CEA-Grenoble, DRFMC / SPSMS / LaTEQS, 17 rue des Martyrs, F-38054 Grenoble cedex 09, France}
\author{F. Lefloch}
\email[Corresponding author: ]{francois.lefloch@cea.fr}
\affiliation{CEA-Grenoble, DRFMC / SPSMS / LaTEQS, 17 rue des Martyrs, F-38054 Grenoble cedex 09, France}
\author{H. Courtois}
\affiliation{Institut N\'eel, CNRS and Universit\'e Joseph Fourier, 25 rue des Martyrs, F-38042 Grenoble cedex 09, France. Also at Institut Universitaire de France}
\author{M. Sanquer}
\affiliation{CEA-Grenoble, DRFMC / SPSMS / LaTEQS, 17 rue des Martyrs, F-38054 Grenoble cedex 09, France}

\begin{abstract}
We present current noise measurements in a long diffusive superconductor - normal metal - superconductor junction in the low voltage regime, in which transport can be partially described in terms of coherent multiple Andreev reflections. We show that, when decreasing voltage, the current noise exhibits a strong divergence together with a broad peak. We ascribe this peak to the mixing between the ac-Josephson current and the noise of the junction itself. We show that the junction noise corresponds to the thermal noise of a non-linear resistor $4k_BT/R$ with $R=V/I(V)$ and no adjustable parameters. 
\end{abstract}


\maketitle

The charge transport in a diffusive superconductor - normal metal - superconductor (S-N-S) junction occurs mainly through multiple Andreev reflections (MAR) \cite{Octavio83}: normal metal quasiparticles with an energy $\epsilon < \Delta$ go through successive Andreev reflections at the two N-S interfaces until they reach the superconducting gap energy $\Delta$. In one Andreev reflection, an incident electron (hole) is retro-reflected as a hole (electron). With an applied bias $V$, the number of successive MAR is equal to $2\Delta / eV +1 $. A case of particular interest deals with long junctions defined by a junction length $L$ larger than the superconducting coherence length $\xi_S$ but still smaller than the single particle coherence length $L_{\phi}$, in contrast with the short junction case $L < \xi_S$ \cite{Naveh99} met for instance in superconducting atomic point-contacts \cite{Cuevas99, Cron01}.

In a long S-N-S junction, the Thouless energy $E_{Th}=\hbar D /L^2$ ($D$ being the electron diffusion constant) is smaller than the gap $\Delta$ and sets the energy scale for the coherent transport. Thus, two regimes can be met depending on the voltage bias. In the (high-bias) incoherent regime $eV > E_{Th}$, the electron and the hole have, away from the interface, independent trajectories. The electronic transport then occurs by incoherent MAR, which induce excess noise as compared to a similar N-N-N system. The experimentally measured shot noise \cite{Hoss00, Hoffmann04} can be described in the framework of the semiclassical theory \cite{Bezuglyi01, Nagaev01, Pilgram05}.

In the (low-bias) coherent regime $eV<E_{Th}$ of interest here, coherent multiple Andreev reflections occur within an energy window of width $E_{Th}$. Since the quasi-particles need to overcome the gap before leaving the normal metal, other processes like incoherent MAR and inelastic interactions are relevant. This makes the description of the coherent electronic transport in a S-N-S junction rather rich and complex. While the current-voltage (I-V) characteristics can be calculated at zero temperature \cite{Cuevas06}, the noise has not yet been derived. Recent measurements \cite{Hoss00, Hoffmann04} show that the noise in such junctions is strongly enhanced at low voltage, which is partially described in the framework of coherent MAR. However, this analysis leads to surprisingly large effective charges compared to the expected cut-off of coherent MAR due to inelastic processes.

Thus the coherent transport in a long S-N-S junction is poorly understood yet. In particular, such a junction is a strongly non-linear conductor, where the differential resistance $dV/dI$ is bias-dependent and generally differs greatly from the normal-state resistance $R_N$ or the ratio $V/I$. The non-linearity appears on a voltage scale $eV$ much smaller than the thermal energy $k_BT$. In the framework of the fluctuation - dissipation theorem, this raises the question of what dissipation-related quantity should be taken into account. 

In this letter, we focus on the shot noise in the coherent regime of long diffusive S-N-S junctions ($eV < E_{Th}<\Delta$). Our low temperature noise measurements show that, when decreasing bias, the current noise exhibits a strong divergence together with a broad peak. We identify this peak as due to the non-linear mixing of the ac-Josephson current at the Josephson frequency $\omega_J /2\pi = 2eV/h$ with the noise of the junction itself. We can then extract, from the measured noise, the intrinsic noise of the junction. The central result of this paper is that the diverging low-bias noise corresponds exactly to the thermal current noise of a non linear resistor with $R=V/I$, which throws new light on the problem of the fluctuation - dissipation theorem in non-linear conductors. 

\begin{figure}
\includegraphics[width=8.5cm]{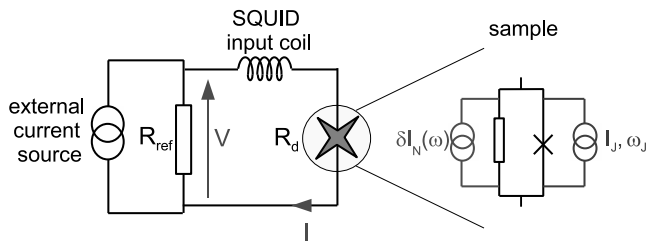}
\caption{Left: Scheme of the measurement circuit. The mean current in the sample and the current fluctuations are measured through the input coil of the SQUID. The reference resistor $R_{\rm ref}$ is at the same temperature than the sample of differential resistance $R_d$. Right: Scheme of the sample, considered as a mesoscopic conductor with a fluctuations source $\delta I_N (\omega)$ in parallel with a Josephson junction oscillating at $\omega_J$. } 
\label{schema}
\end{figure}

We measured the transport properties and the current noise of S-N-S junctions using a superconducting quantum interference device (SQUID) experimental set-up \cite{Jehl99} inserted in a dilution fridge ($T_{\rm min}=30$ mK). The SQUID is connected to a resistance bridge (Fig. \ref{schema}) composed of the sample on one side and of a reference resistor $R_{\rm ref}$ on the other side. The current fluctuations propagating in this loop come from the reference resistor and from the sample. The value of the reference resistor $R_{\rm ref}=29$ m$\Omega$ is chosen so that most of the measured current noise comes from the sample. The intrinsic noise level is about 8 $\mu \Phi_0$/$\sqrt{{\rm Hz}}$, which corresponds to 1.6~pA/$\sqrt{{\rm Hz}}$ in the SQUID input coil. Hereafter, the current noise measurements were performed in the frequency range between 4~kHz and 6.4~kHz.

\begin{figure}
\includegraphics[width=8cm]{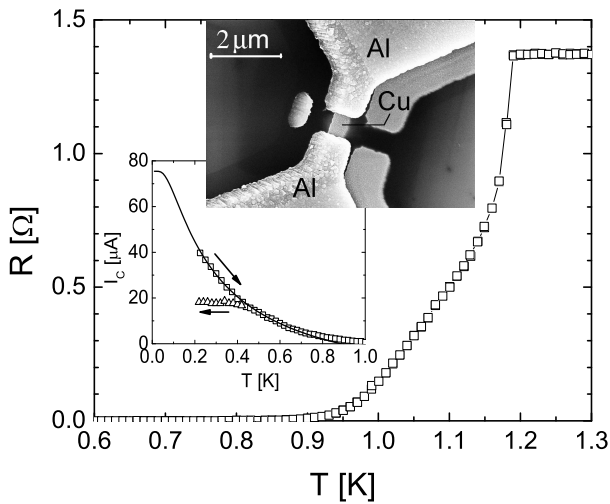}
\caption{Zero bias resistance $R_{d}=dV/dI$ \textit{vs.} temperature $T$ measured with $I_{ac}=25$ nA. Top inset: scanning electron micrograph of the sample. Bottom inset: Critical current $I_C$ (square symbols) and retrapping current $I_R$ (triangles) \textit{vs.} $T$. The solid line is a fit (see text for equation).} 
\label{figRT}
\end{figure}

The sample is a Cu bridge of 700~nm length and 425~nm width connected to two thick superconducting Al electrodes (see Fig.~\ref{figRT} inset). It has  been fabricated by two-angles evaporation through a PMMA-PMMA/MAA bilayer mask in a ultra-high vacuum chamber. The Cu and Al thicknesses are 50 and 500 nm respectively. From the Cu diffusion constant estimated to $D=100$ cm$^2$.s$^{-1}$ by resistance measurements, we get the Thouless energy of the sample $E_{Th}$ equal to 13 $\mu$eV.  

The zero bias resistance is represented as a function of temperature in Fig. \ref{figRT}. It shows a small drop at 1.18 K due to the transition of the Al reservoirs. At lower temperatures, the resistance decreases slowly and goes to zero at about 850 mK. Below this temperature, a supercurrent is established. We measured the critical current $I_C$ as a function of temperature down to 200 mK (see inset of Fig. \ref{figRT}). It follows the usual behavior observed in long diffusive S-N-S junctions and can be fitted using $eR_N I_C=aE_{Th}\left(1-b \exp \left(-aE_{Th}/3.2k_BT \right) \right)$. As for parameters, we took $R_N=0.85\ \Omega$ and $a \simeq 5 $ and set $b=1.3$ \cite{Dubos02}. Such a value for $a$ shows that the possible barrier resistance is negligible. Below 450~mK, the IV characteristic is hysteretic, with a retrapping current $I_R$ significantly below the critical current. In the following, we will focus on the non-hysteretic regime, $T \geq 500$ mK.  

\begin{figure}
\includegraphics[width=7.5cm]{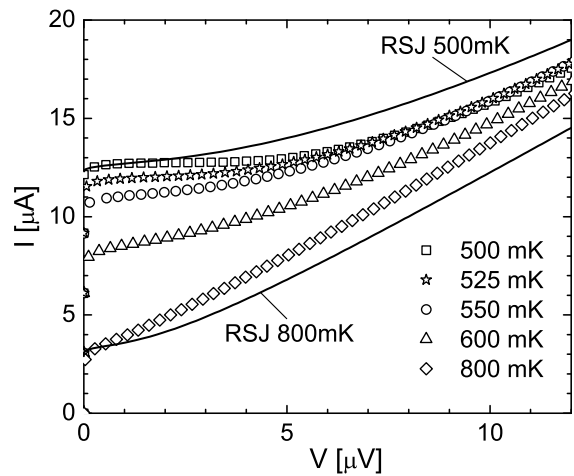}
\caption{IV curves at several temperatures between 500 and 800 mK. The lines are obtained from the RSJ model at 500 and 800 mK \cite{Bishop78}, with the parameters: $R_N=0.85\ \Omega$ and the critical currents set to 12.7 and 3.4 $\mu$A respectively. } 
\label{figIV}
\end{figure}

IV characteristics measured in this regime are shown in Fig.~\ref{figIV} together with the resistively shunted junctions (RSJ) calculation at finite temperature \cite{Bishop78} for $T=500$ and 800 mK. As already mentionned in other studies on S-N-S junctions \cite{Hoss00,Dubos01}, the curves cannot be described within the RSJ model. Because all the experimental curves are lying between the 500 mK and 800 mK theoretical curves, the discrepancy cannot be simply due to heating effect. From the overall shape of IV characteristics, we can state that some current in the junction is supported by a normal "channel" (see right of Fig. \ref{schema}), since in pure Josephson junctions no current can flow at a voltage less than the gap. 

\begin{figure}
\includegraphics[width=8cm]{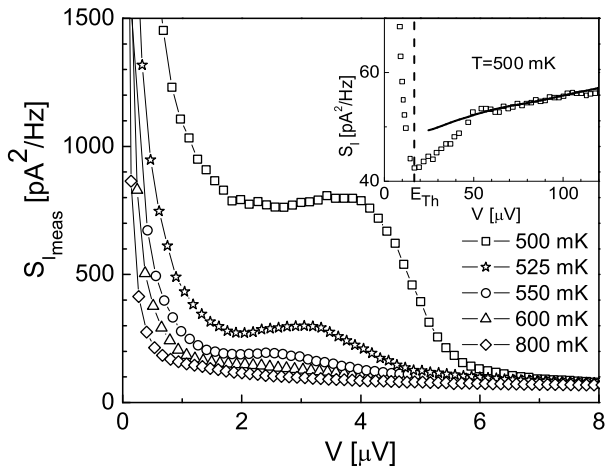}
\caption{Current noise $S_{I_{\rm meas}}$ \textit{vs.} bias voltage $V$ at several temperatures between 500 and 800 mK. Inset: $S_I$ at 500 mK up to 120~$\mu$V. The solid line is the noise prediction due to incoherent MAR \cite{Hoffmann04, Bezuglyi01, Nagaev01}.}
\label{figSV}
\end{figure}

We first measured the shot noise over a large voltage scale (See Fig.~\ref{figSV} inset). The voltage dependence at $V>50\ \mu$V can be nicely described within the semiclassical theory of incoherent MAR \cite{Bezuglyi01, Nagaev01} (with $\Delta = 163\ \mu$eV) as already reported in a previous work on the same kind of junctions \cite{Hoffmann04}. We also confirmed the existence of a noise minimum at the Thouless energy, which appears as a signature of the transition towards the coherent regime at low voltage, on which we will focus now. 

Fig. \ref{figSV} shows the low voltage ($V<8\ \mu$V) part of the noise spectral density as a function of bias, at five temperatures from 500 to 800~mK. The first striking result is the overall divergence of the noise at low bias for every temperature. The second important feature is the broad peak with a maximum between 2 and 4 $\mu$V. When the temperature increases, it is less and less pronounced, moves to lower voltages and finally disappears around 800 mK. 
We analyze this broad peak in terms of a mixing between the ac-Josephson current $I_J$ and the fluctuating current $\delta I_N$ generated by the junction through the non-linear response of the junction. In such a situation, the voltage fluctuations $S(\omega)$ at low frequency are determined not only by the current fluctuations at the frequency $\omega$, but also by those close to the Josephson frequency $\omega_J$ and its harmonics, inducing the so-called mixed down noise. Following Likharev and Semenov \cite{Likharev72}, the resulting noise spectral density is:
\begin{equation}
S_V(\omega )= \sum_k | Z_k| ^2 S_I(k\omega_J-\omega) 
\label{EquLikharev1}
\end{equation}
where $S_I$ is the current noise spectral density of the fluctuations source and $Z_k$ are the Fourier coefficients of the junction impedance ($V(\omega)=\sum_k Z_k I(\omega - k \omega_J)$). 
Note that, due to the large SQUID impedance at the Josephson frequency $\omega_J$, there is no coupling between the S-N-S junction and the reference resistor at $\omega_J$. Thus, the reference resistor can not contribute to the mixed-down noise.

At low frequency ($\omega \ll \omega_J$), only the first harmonics ($k=-1,0,1$) of the ac-Josephson current are relevant \cite{Koch80}. Therefore, Eq.~(\ref{EquLikharev1}) becomes at zero frequency $S_V(0)= R_d^2 S_I(0)+2 | Z_1| ^2 S_I(\omega_J)$ where $R_d$ is the differential resistance $\partial V / \partial I$. For an arbitrary current-phase relation, Kogan and Nagaev \cite{Kogan88} have shown that the coefficient $| Z_1| ^2$ is equal to $-V/4 \times \partial R_d / \partial I$. This leads to : 
\begin{equation}
\begin{gathered}
S_V(0)= R_d^2 [S_I(0)-\eta S_I(\omega_J) ] 
 \, {\rm with} \quad \eta = \frac{V}{2R_d^2}\frac{\partial R_d}{\partial I}
\label{EquNagaev} 
\end {gathered}
\end{equation}

\begin{figure}
\includegraphics[width=8.5cm]{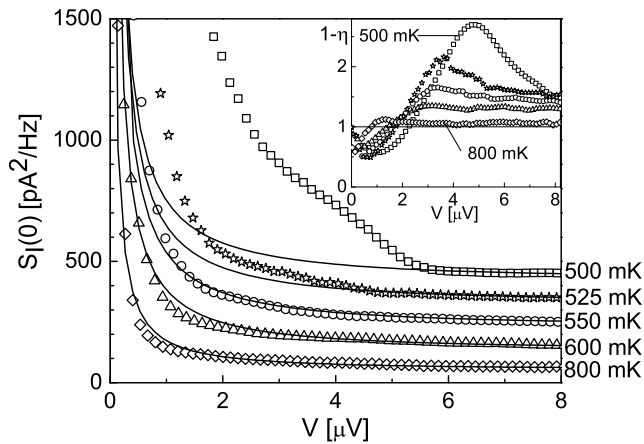}
\caption{Current noise $S_I(0)=S_{I_{\rm meas}}/(1-\eta)$ \textit{vs.} bias voltage $V$ at several temperatures between 500 and 800 mK (Data are sucessively shifted by 100 pA$^2$/Hz for clarity). Lines are the calculated noise $4k_BT/(V/I)$. Inset: $1-\eta$ \textit{vs.} $V$ at the same temperatures, where $\eta$ is the coupling factor (See Eq. (\ref{EquNagaev})).}
\label{figSVfit}
\end{figure}

In our experimental set-up, the SQUID measures the resulting low frequency current noise that propagates through the SQUID input coil. It is given by 
$S_{I_{\rm meas}}(0)= S_V(0)/R_d^2=S_I(0)-\eta S_I(\omega_J)$. Because the thermal energy $k_B T$ is larger than the Josephson frequency $\omega_J /2\pi = 2eV/h$, we can assume that the spectral noise $S_I(\omega_J)$ of the junction does not differ from its zero-frequency value $S_I(0)$. Then,
\begin{equation}
S_{I_{\rm meas}}(0)= S_I(0)(1-\eta )
\label{EquMeasure2}
\end{equation}
We obtained the coupling factor $\eta$ from the measured quantities $V$ and $R_d$ by numerically differentiating $R_d$ with respect to the calculated $I$. The quantity $1-\eta$ is plotted in the inset of Fig. \ref{figSVfit}. We can then extract, from the measured noise, the intrinsic junction noise $S_I(0)=S_{I_{\rm meas}}(0)/(1-\eta)$, see Fig. \ref{figSVfit}. For all the curves the broad peak has disappeared or, at least, has been strongly reduced \cite{footnote1}. This clearly justifies the mixed down noise analysis using the generalized coupling factor $\eta$. To our knowledge, this prediction of Kogan and Nagaev \cite{Kogan88} had never been verified experimentally.

From Fig. \ref{figSVfit}, we see that $S_I(0)$ exhibits a strong divergence at low voltage. Therefore, it cannot be described by the thermal noise of the normal resistance $4 k_BT /R_N$ as it was considered in Ref. \onlinecite{Decker75} and \onlinecite{Koch82}. We cannot either understand the data by introducing an effective temperature. Indeed, we would get an effective temperature as high as 100~K which appears unrealistic since the aluminum critical temperature is 1.2~K. 

We propose, in our temperature and voltage regime, that the noise of the junction can be written as: 
\begin{equation}  
S_{I}(0)=\frac{4k_BT}{V/I}
\label{EquSn}
\end{equation}
The comparison with experimental data of $S_I(0)$ is shown in Fig.~\ref{figSVfit} with {\it no adjustable parameter}. Between 800 and 550 mK, the agreement is remarkable. 

The use of Eq. (\ref{EquSn}) can be understood in two ways.  First, it is the low voltage limit ($eV \ll k_BT$) of the quasiparticle noise of a junction with low transmitting channels : $S_{I}(0)=2eI \coth \left(eV/2k_BT \right)$ \cite{Dahm69, footnote2}. If effective charges $e^*>e$ are responsible for the transport, $e$ should be replaced by $e^*=Ne$ in the above expressions. In that case, the condition $e^*V=NeV \ll k_BT$ for Eq. \ref{EquSn} to remain valid, is verified as long as  $N < 30$ at $1\ \mu$V and 550 mK. In this description, only channels with low transmission coefficients contribute to the current and to the noise. 
Eq.~(\ref{EquSn}) might also be related to the fluctuations-dissipation theorem in a non-linear system, in which the resistance $R_N$ would be replaced by $V/I(V)$. Here, dissipation arises from electron-electron interactions that necessarily occur at low voltage due to the large number of Andreev reflections needed for the quasiparticles to escape into the superconducting electrodes. In both cases, we conclude that, at very low voltage and "high" temperature, the current noise in long S-N-S junctions is governed by the transport of incoherent quasiparticles between the two superconducting reservoirs, which induces a noise divergence when the voltage goes to zero. 

When decreasing further the temperature, the experimental data present a large excess noise compared to the noise obtained from Eq.~(\ref{EquSn}). Therefore, additional noise sources must be involved. It is worth noting that this deviation from a pure quasiparticles noise occurs below 525 mK, which corresponds to $k_BT \simeq 3.5 E_{Th}$, very close to the mini-gap width ($3.1 E_{Th}$) \cite{Zhou98}. This suggests that coherent processes, that become relevant in that energy scale, should play a role here. 

In conclusion, we have presented current noise measurements in the non-hysteretic temperature range of a long diffusive S-N-S junction at very low bias ($eV<E_{Th}<\Delta$). The observed noise spectral density is enhanced by several orders of magnitude compared to the thermal noise of a normal junction. It exhibits a broad peak arising from the non-linear coupling between the junction current noise and the ac-Josephson current. This mechanism is of particular interest for studying the high frequency (quantum) noise of a dissipative conductor by means of low frequency measurements \cite{Koch82}. Here the dissipative channel is perfectly coupled to the non-linear oscillator since they are both part of the same junction. The intrinsic noise of the junction that we could extract corresponds extremely well to the thermal-like noise $4k_BT/R$ of a non-linear resistor $R=V/I(V)$. This shows up as a divergence at low voltage and appears to be analogous to the quasiparticles noise of a normal junction with channels of low transmission coefficients. 

We acknowledge M. Houzet and F. Hekking for fruitful and continuous discussions. Samples have been fabricated at Nanofab-CNRS Grenoble. This work was funded by EU STREP project 'SFINx' and the IPMC/Grenoble.


\begin{thebibliography}
\bibitem{footnote1} At 500 mK, the differential resistance $R_d$ is so large ($>35\ \Omega$, compared to $R_N=0.85\ \Omega$), that the experimental set-up is not well adapted for such measurements. This could induce an error in the determination of $\eta$.
\end{thebibliography}
\end{document}